\documentclass[12pt]{iopart}
\usepackage[utf8]{inputenc}
\usepackage{amssymb,amsfonts,setspace,graphicx,bm,float,amssymb,siunitx}

\newcommand{\bra}[1]{\langle #1|}
\newcommand{\openone}{\leavevmode\hbox{\small1\kern-2.8pt\normalsize1}}
\newcommand{\ket}[1]{|#1\rangle}

\expandafter\let\csname equation*\endcsname\relax

\expandafter\let\csname endequation*\endcsname\relax
\usepackage{amsmath}

\begin{document}
\title{Logical measurement-based quantum computation in circuit-QED}

\date{\today}
\author{Jaewoo Joo}
\address{School of Computational Sciences, Korea Institute for Advanced Study, Seoul 02455, Korea}
\address{Clarendon Laboratory, University of Oxford, Parks Road, Oxford OX1 3PU, UK}

\author{Chang-Woo Lee} 
\address{School of Computational Sciences, Korea Institute for Advanced Study, Seoul 02455, Korea}

\author{Shingo Kono}
\address{Research Center for Advanced Science and Technology (RCAST), The University of Tokyo, Meguro-ku, Tokyo 153-8904, Japan}

%\author{Yasunobu Nakamura}
%\address{Research Center for Advanced Science and Technology (RCAST), The University of Tokyo, Meguro-ku, Tokyo 153-8904, Japan}
%\address{Center for Emergnent Matter Science (CEMS), RIKEN, Wako, Saitama 351-0198, Japan}

\author{Jaewan Kim} 
\address{School of Computational Sciences, Korea Institute for Advanced Study, Seoul 02455, Korea}

\begin{abstract}
We propose a new scheme of measurement-based quantum computation (MBQC) using an error-correcting code against photon-loss in circuit quantum electrodynamics. We describe a specific protocol of logical single-qubit gates given by sequential cavity measurements for logical MBQC and a generalised Schr\"odinger cat state is used for a continuous-variable (CV) logical qubit captured in a microwave cavity. It is assumed that a three CV-qudit entangled state is initially prepared in three jointed cavities and the microwave qudit states are individually controlled, operated, and measured through a readout resonator coupled with an ancillary superconducting qubit. We then examine a practical approach of how to create the CV-qudit cluster state via a cross-Kerr interaction induced by intermediary superconducting qubits between neighbouring cavities under the Jaynes-Cummings Hamiltonian. This approach could be scalable for building 2D logical cluster states and therefore will pave a new pathway of logical MBQC in superconducting circuits toward fault-tolerant quantum computing.
\end{abstract}

\maketitle
\section{Introduction}

Measurement-based quantum computation (MBQC) offers a new platform of quantum information (QI) processing. Quantum algorithms are performed by sequential single-qubit measurements in multipartite entangled states initially (e.g., cluster states \cite{Cluster}) instead of massive controls of individual qubits during the whole information processing \cite{MBQC1,MBQC2}. This advantage is, however, only beneficial for QI processing if the specific multipartite entangled state can be initially well-prepared and the capability of fast and precise single-qubit measurements are viable. For example, a two-qubit cluster state is the simplest resource state for MBQC given by $\ket{2CS}_{AB} = (\ket{0}_A \ket{+}_B + \ket{1}_A \ket{-}_B)/\sqrt{2}$ with $\ket{ \pm} = (\ket{0} \pm \ket{1})/\sqrt{2}$.
If the operation angle $\theta$ is chosen for the measurement basis vectors in qubit $A$, $\ket{ \pm \theta} = (\ket{0} \pm e^{-i \theta} \ket{1})/\sqrt{2}$, the resultant state in $B$ after the measurement $\ket{\pm \theta} \bra{\pm \theta}$ on $A$ becomes a single-qubit operated state such as ${}_{A} \bra{ \pm \theta} 2CS \rangle_{AB} \propto \hat{e}^{\pm} e^{\pm i {\theta \over 2}}  H R^z (\theta) \ket{+}_{B}$, for Hadamard gate $H= ( X + Z )/\sqrt{2}$, $z$-axis rotation operator $R^z(\theta) = e^{-i{\theta  \over 2}  } \ket{0}\bra{0} + e^{i{\theta  \over 2} } \ket{1}\bra{1} $ and $\hat{e}^{\pm}= \{ \openone, X\}$ with Pauli operators $X, Z$. Thus, it is interpreted as the single-qubit gate $H R^z (\theta)$ is performed on $\ket{+}$ by the measurement of qubit $A$ with the chosen angle $\theta$ on $\ket{2CS}_{AB}$. It is therefore of essence to demonstrate efficiently building such a useful entangled resource state and performing single-qubit measurements on the resource state for practical MBQC.

The MBQC in continuous variables (CVs) has been firstly well developed in quantum optics since such CV cluster states are achievable using traveling squeezed states through optical parametric amplifiers \cite{CV-MBQC1, Sam_Braunstein1, Sam_Braunstein2, Sam_Braunstein3}. For example, the recent development of creating one-dimensional (1D) and 2D CV cluster states has been demonstrated in quantum optics using quantum memory and in time/frequency domain \cite{CV-MBQC_ex1,CV-MBQC_ex2}. In these methods, a phase-space translation operator is in general used for single-qubit gates while a two-qubit controlled-Z gate is implemented in a sequence of beam-splitters \cite{Furusawa2011}. Toward fault-tolerant CV MBQC using this approach, a scheme of high squeezing photons (20.5 dB) has been required to reach the error tolerance threshold with $10^{-6}$ through concatenated codes \cite{FT_CV_MBQC}, and is very challenging with the state-of-the-art experiments in quantum optics. Recently, an alternative method of creating four-qubit CV cluster states has been suggested in a circuit quantum electrodynamics (circuit-QED \cite{BlaisPRA}) system \cite{CV-cluster-CQED}.

One of the advantages of using CVs is that the optical cluster states are built in a deterministic manner and can in principle store information in infinite dimension \cite{GKP, Jeong02-1, CVentanglement, New_archive} while alternative optical methods of creating discrete-variable cluster states have been in general generated in polarization or spatial modes probabilistically by using parametric down conversion \cite{DVPhotonicMBQC}. We will in particular use a specific logical qubit encoded in generalised Schr\"odinger cat states, which are the superposition of phase-rotated coherent states \cite{Schro01}. 
It is known that the specific CV-qudit states can be used for the error-correctable QI unit against particle-loss and have been successfully demonstrated in circuit-QED for practical quantum memory \cite{qcMAP13,NewYale16,NewYale17,Yale_FTcat}. This circuit-QED approach could thus be advantageous for error-correctable quantum computing equipped with photon-loss resilience in the CV-qudit code \cite{catcode, Yale-QECC}.

We here propose a novel circuit-QED scheme of performing logical qubit gates and the desired outcome is achieved by cavity measurements from a tripartite CV-qudit cluster state as a single-qubit operated state in the CV-qudit code. Because it might be concerned how to initially implement the complex multipartite cluster state by the manual controls of cavity states, we first suggest a circuit-QED architecture capable of building the target CV-qudit entangled state using an induced cross-Kerr interaction, which naturally provides an entangling gate between neighbouring cavity qudits. It is known that one can in principle engineer cross-Kerr interaction in the multiple-cavity architecture with tunable self-Kerr interaction \cite{Matt17}. Then, after we define the CV-qudit and its cluster states, we present a new protocol for a logical single-qubit gate in MBQC using three specific circuit-QED techniques such as a coherent-state measurement, parity measurement, and a selective number-dependent arbitrary phase (SNAP) gate. All these techniques have been well developed and demonstrated in theory and experiment \cite{1-photon_Kerr, EranPRB, SNAPgate}. We finally examine the cross-Kerr entangling scheme of builiding two CV-qudit cluster states with an intermediary superconducting qubit and this circuit-QED architecture would enable to investigate not only QI processing but also more broader sciences including many-body physics \cite{manybody} and quantum chemistry \cite{quan_chem} in the future. 

\section{Results}
\begin{figure}[t]
\includegraphics[width=9cm,trim=3cm 5.5cm 3cm 0.5cm]{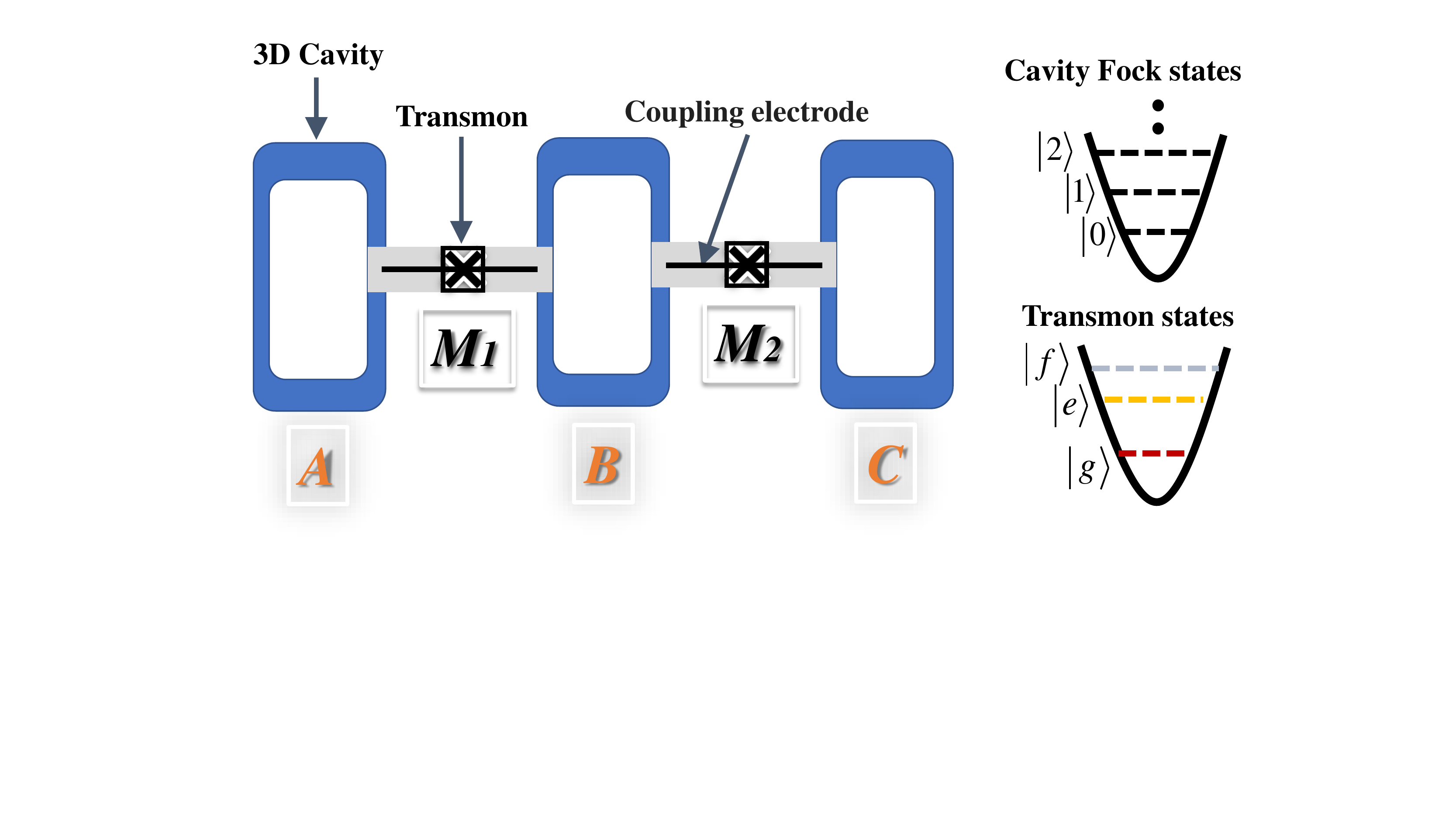}
\includegraphics[width=7.5cm,trim=0.5cm 6cm 0cm 0.5cm]{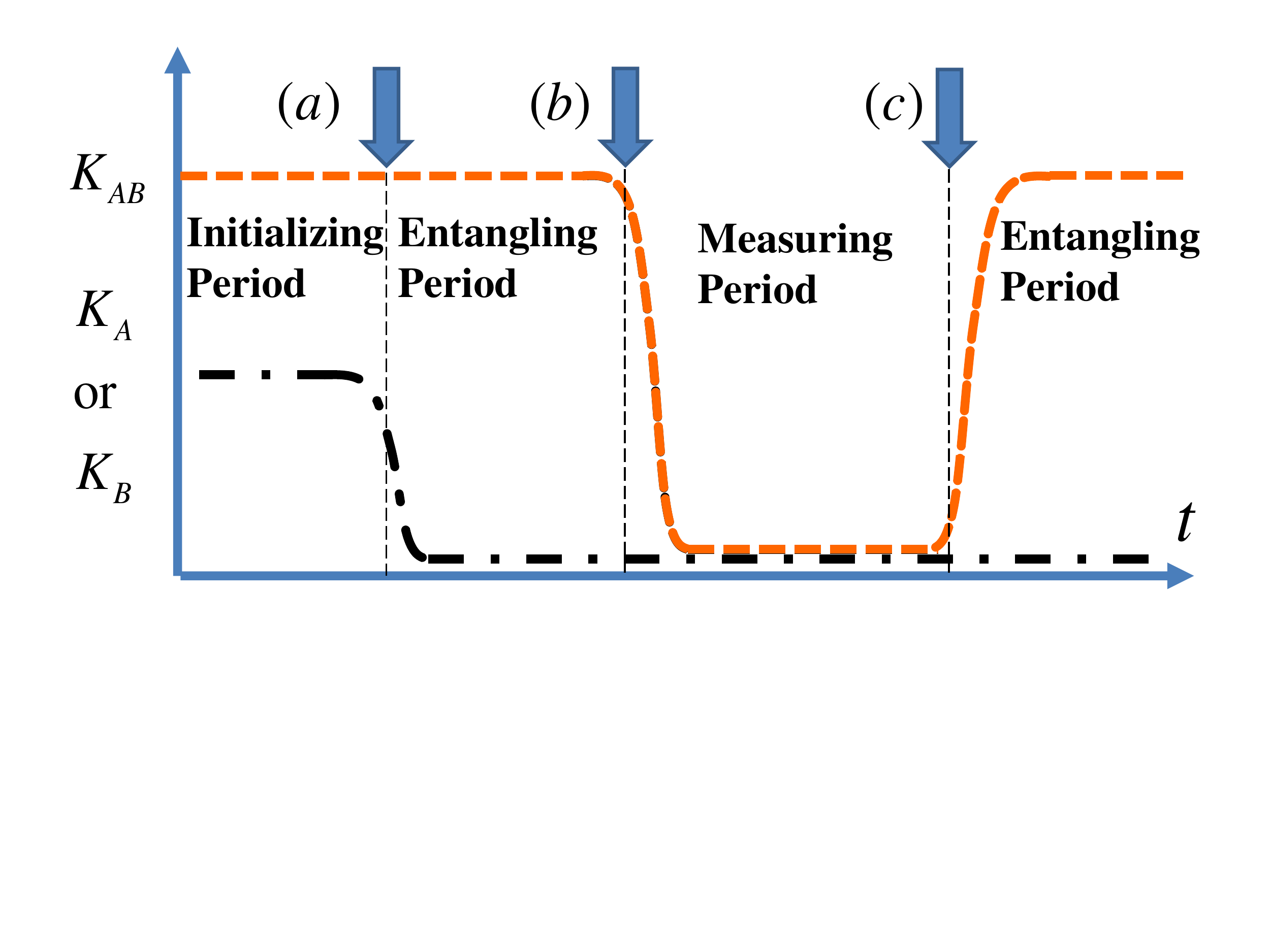}
\caption{Schematics of logical MBQC in a circuit-QED architecture.
(Left) Three cavities ($A, B, C$) have the intersected superconducting qubits $M_1$ and $M_2$ used for inducing the Kerr interactions between cavities. When a 3-qudit logical cluster state is built in the cavities by cross-Ker interaction ($K_{ij}$), logical MBQC is performed by a sequential measurement of each cavity. 
The colours of transmon\rq{}s energy states represent the anharmonicity of the energy levels in a transmon. (Right) the tunability of Kerr effects between the neighbouring cavities provided with the help of tunable on-site superconducting qubits and an extra (tunable) intermediary qubit in the same architecture (the details are shown in \cite{Joo_SR,Matt17}). For example, the self-Kerr effects can be only reduced by shifting energy levels in on-site qubits at point (a) and the simultaneous entangling gates are performed by cross-Kerr $K_{ij}$ between (a) and (b). From (b) to (c), the cavities are uncoupled and the sequential measurements of each cavity are performed for MBQC.
 \label{fig:02} }
\end{figure}

\subsection{Circuit-QED architecture for entangled cavity states}
The platform of superconducting circuits has been rapidly developed for QI processing over two decades \cite{Nakanmura}. The artificial qubits are intrinsically scalable and manufacturable in the forms of different qubit types with precise control of desired parameters \cite{transmon, flux}. In experiment, one utilises only superconducting qubits (mainly transmon qubits \cite{transmon}) for QI unit while it has also been successfully shown that a coupled system of superconducting qubits and 3D cavities offers excellent capability of creating quantum cavity states through the nonilnearity of an intermediary superconducting qubit, e.g., deterministic generation of Schr\"odinger cat states and entangling CV states inside the cavities \cite{Yale_teleport, Yale_big_cat}.

As shown in the left figure of Fig.~\ref{fig:02}, we consider a circuit-QED architecture for creating entangled microwave states and the neighboring cavities are connected with each other via a middle transmon qubit $M_i$ ($i=1,2$) enabling to entangle cavity states. This approach shows a unique advantage that a massive 1D CV-qudit cluster state can be built in one step as the key resource state for MBQC. Since two cavities are simple harmonic oscillators, a superconducting qubit inserted in between two cavities brings induced Kerr effects on the joint cavity modes. For an ideal case, it is assumed that two neighboring cavities are only coupled by a cross-Kerr interaction, which is induced by the intermediary superconducting qubit. 

In a real circuit-QED setup, this architecture might cause unwanted nonlinear effects over the cavities (e.g., self-Kerr distortion effects and non-identical cross-Kerr effects). In general, the cavity self-Kerr effect makes the amount of distortion in the cavity state and could prevent building ideal CV-qudit entangled states and to measure the cavity qubit accurately at an appropriate time. 
For example, let us consider the Jaynes-Cummings (JC) Hamiltonian for two cavities with an intermediary transmon is given by  
\begin{eqnarray}
\label{total_Ham01}
\hspace{-1cm} \hat{H}^{JC}_{ABM} &&= \sum_{c=A,B,M}  \omega_c \hat{a}^{\dag}_{c} \hat{a}_{c}  + {K_M} \hat{a}^{\dag}_{M} \hat{a}_{M} \hat{a}^{\dag}_{M} \hat{a}_{M}  +  \sum_{c=A,B} \lambda^{M}_{c} (  \hat{a}^{\dag}_{M} \hat{a}_{c} + \hat{a}_{M} \hat{a}^{\dag}_{c}),
\end{eqnarray}
with creation operator $\hat{a}^{\dag}$ and $\hbar =1$.
It is experimentally confirmed that self- and cross-Kerr effects exist in the cavities coupled with a superconducting qubit \cite{1-photon_Kerr,SNAPgate} and theoretically the adiabatic elimination theory can show the existence of these effects (upto the fourth order in the JC Hamiltonian \cite{Matt17,AdiabaticRef}). We will examine the validity of the induced cross-Kerr interaction in this architecture to build a two CV-qudit cluster state in Section \ref{Implement_s02}. 

Fortunately, a Kerr-engineering scheme has been recently proposed to amend self- and cross-Kerr effects in a qubit-cavity array and is applicable for creating a desired 1D CV-qudit entangled state with the help of extra tunable superconducting qubits in a similar architecture \cite{Matt17}. For example, suppose that a flux qubit is additionally attached on each cavity. In Ref.~\cite{Matt17}, it is shown that the controls of energy levels of the flux qubit diminish the amount of self-Kerr interaction $K_j$ in each cavity, but the cross-Kerr interaction still survives between neighbouring cavities. As shown in the right figure of Fig.~\ref{fig:02}, two cavity states starts to be entangled with $K_j \approx 0$ during the period between $(a)$ and $(b)$. After the entangling period, the cross-Kerr interaction can be also reduced in a similar technique and the cavity states can be effectively decoupled for better performance of individual cavity measurements between $(b)$ and $(c)$ (see Fig.~5 in  \cite{Matt17}). For logical MBQC, we need to perform a type of quantum-non-demolition (QND) measurements on each cavity and their details are addressed in Section \ref{Sec:operation}.

\subsection{Cat qudits}
\label{2-1}
We first introduce the definition of CV qudits (with $d=4$) written in the superposition of phase-encoded coherent states. The CV qudits are defined by
 \begin{eqnarray}
\label{Simple_03-1}
\ket{{0}_4} && =  M^{0}_{\alpha} \left( \ket{\alpha} + \ket{i \alpha} + \ket{-\alpha} + \ket{-i \alpha} \right) = \sum_{m=0}^{\infty} c_{0m} \ket{4m}, \\
\label{Simple_03-3}
\ket{{1}_4} && =  M^{1}_{\alpha} \left( \ket{\alpha} - i \ket{i \alpha} - \ket{-\alpha}+ i \ket{-i \alpha} \right) = \sum_{m=0}^{\infty} c_{1m} \ket{4m+1},  \\
\label{Simple_03-4}
\ket{{2}_4} && =  M^{2}_{\alpha} \left( \ket{\alpha} - \ket{i \alpha} + \ket{-\alpha} - \ket{-i \alpha} \right)  = \sum_{m=0}^{\infty} c_{2m} \ket{4m+2},  \\
\label{Simple_03-2}
\ket{{3}_4} && =  M^{3}_{\alpha} \left( \ket{\alpha} +i \ket{i \alpha} - \ket{-\alpha} - i \ket{-i \alpha} \right) = \sum_{m=0}^{\infty} c_{3m} \ket{4m+3},  
 \end{eqnarray}
where a coherent state with real values $\alpha$ and $\phi$ is $\ket{\alpha e^{i\phi}} = e^{-|\alpha|^2/2} \sum_{n=0}^{\infty} {\alpha^n e^{i \phi n} \over \sqrt{n!}} \ket{n}$ and $\ket{4m+j}$ is a Fock state with $4m+j$ photons ($M_{\alpha}$ as a normalisation factor).
Note that their complementary qudits are defined as $\ket{\tilde{0}_4} = \ket{\alpha}$, $\ket{\tilde{1}_4} = \ket{i \alpha}$, $\ket{\tilde{2}_4} = \ket{-\alpha}$, and $\ket{\tilde{3}_4} = \ket{-i \alpha}$ \cite{Kim15}. 

The generalised Pauli operators for the qudits are defined by $\hat{Z}_4\ket{\tilde{k}_4} = \ket{\widetilde{(k+1)}_4}$ and $\hat{X}_4 \ket{{(k+1)}_4} = \ket{{k}_4}$. The qudit Pauli operators can be physically implemented by phase rotation $\hat{Z}_4 = e^{ i {\pi \over 2} (\hat{a}^{\dag} \hat{a}) }$ and photon addition $\hat{X}_4 \approx \hat{a}^{\dag} / \sqrt{ \langle \hat{a}^{\dag}\hat{a} \rangle }$ (or photon subtraction $\hat{a} / \sqrt{ \langle \hat{a}^{\dag}\hat{a} \rangle }$). Note that the normalisation coefficients $M^{i}_{\alpha}$ are approximately equal to 1/2 for $\alpha \ge 2$, which implies the validity of orthogonality in qudit $\ket{k_4}$ for QI unit ($k=1,2,3,4$). In other words, if the average photon number should be large enough to distinguish between coherent states, the qudits can be used for logical qubits against photon-loss errors in Section \ref{sec:LogicalQubit} \cite{catcode}.

\subsection{How to create ideal three CV-qudit cluster states}
We here show an mathematical description of how to build 1D CV-qudit cluster states with an ideal cross-Kerr interaction \cite{Sanders12,Sanders92}. The cross-Kerr interaction shows a natural way to entangle two coherent states (see the details in Section \ref{Sec:Entangle-two-cat}). For a three-cavity case, an initial state $\ket{\psi^{int}}_{ABC}$ is prepared in three cavities and a time-evolved state at time $t$ is given by 
\begin{eqnarray}
\label{total_Ham00}
\ket{\psi (t)}_{ABC} = \exp \left( i \hat{H}_{ABC}  \, t \right) \ket{\psi^{int}}_{ABC}.
\end{eqnarray}
The cross-Kerr Hamiltonian is ideally given in
$\hat{H}^{tot}_{ABC} = K_{AB} (\hat{a}^{\dag}_A \hat{a}_A) (\hat{a}^{\dag}_B \hat{a}_B) + K_{BC} (\hat{a}^{\dag}_B \hat{a}_B) (\hat{a}^{\dag}_C \hat{a}_C)$. With the assumption $K_{AB} = K_{BC}$ for simplicity, the three CV-qudit state at a quarter of the revival time is written in
\begin{eqnarray}
\label{3QuditCS01}
\hspace{-2cm} \ket{\psi^{ideal} ( {\tau_r / 4} ) }_{ABC}  =   \exp \left( i  {\tau_r \over 4} \hat{H}^{tot}_{ABC}   \right) \ket{\alpha}_{A} \ket{\alpha}_{B} \ket{\alpha}_{C} =  {1 \over 2}  \sum_{k=0}^3 \ket{\tilde{k}_4}_A \ket{{k}_4}_B  \ket{\tilde{k}_4}_C.
\end{eqnarray}

It could be crucial to match the strength values of two cross-Kerr interactions between neighbouring cavities ($K_{AB} = K_{BC}$) to create the target state in Eq.~(\ref{3QuditCS01}). Otherwise, the cavity state becomes maximally entangled in $A$ and $B$ at a certain time but it does not in $B$ and $C$. In Ref.~\cite{Matt17}, a slight modification of the circuit-QED architecture has been investigated with additional superconducting qubits to control self- and cross-Kerr interactions independently. This modified architecture might thus be beneficial for building a multi-partite entangled state in many cavities at once toward practical MBQC.

\subsection{Three single-qudit gates in cavity states} 
\label{Sec:operation}
For logical MBQC, three specific single CV-qudit operations are required in each cavity such as (1) coherent-state projection $\hat{P}^{Coh}$, (2) parity measurement $\hat{P}^{Par}$, and (3) SNAP phase gates. Note that all the gates have already been demonstrated in a qubit-cavity architecture experimentally. 
In a dispersive regime of the JC Hamiltonian, which is defined by much smaller coupling strength than the difference between cavity and qubit frequencies, it is feasible to perform the projection measurement on Fock states in a cavity-transmon coupled system (see the details in Section \ref{sec:coherent-measurement}). 

To describe the operations, we define an arbitrary CV-qudit state $\ket{\Psi_4}_A$ given in cavity $A$ by 
\begin{eqnarray}
\label{arbit_CVqudit01}
\ket{\Psi_4}_A && = a \ket{{0}_4}_A + b \ket{{1}_4}_A + c \ket{{2}_4}_A + d \ket{{3}_4}_A .
\end{eqnarray}
First, the projection set of a coherent-state is given by $\hat{P}^{Coh} (\alpha) = \{ \ket{\tilde{0}_4}\bra{\tilde{0}_4}, \, \openone -\ket{\tilde{0}_4}\bra{\tilde{0}_4}\}$ and is viable in a microwave cavity coupled with a superconducting qubit and a readout resonator \cite{NewYale16}. For example, $\hat{P}^{Coh} (\alpha)\ket{\Psi_4}_A \approx \ket{\tilde{0}_4} = \ket{\alpha} $ for $\ket{\tilde{0}_4}\bra{\tilde{0}_4}$ and the definitions and details are presented in Section \ref{sec:coherent-measurement}. 

Second, a QND parity measurement of cavity states has been successfully demonstrated  with the assistance of an ancillary superconducting qubit in Ref.~\cite{Exp_parity}. The cavity state is projected on the even- or odd-photon subspace such as $\hat{P}^{Par} (even, odd) = \{ \ket{{0}_4} \bra{{0}_4} + \ket{{2}_4} \bra{{2}_4} ,  \ket{{1}_4} \bra{{1}_4}+ \ket{{3}_4} \bra{{3}_4} \}$ and its parity is imprinted in the state of an ancillary readout qubit. For example, the state $\ket{\Psi_4}_A$ is collapsed by the parity measurement into
$\hat{P}^{Par} (even) \Big( \ket{\Psi_4}_A\ket{g} \Big) \propto a \ket{{0}_4}_A + c \ket{{2}_4}_A $ with the outcome of the qubit state in $\ket{e}$ or 
$\hat{P}^{Par} (odd) \Big( \ket{\Psi_4}_A\ket{g} \Big)  \propto b \ket{{1}_4}_A + d \ket{{3}_4}_A$ with $\ket{g}$. Therefore, the cavity state is projected in either the even- or odd-photon subspace through the parity measurement performed by the readout qubit. (see details in Section \ref{sec:parity}).

Finally, the SNAP gate is essential for performing photon-phase operations for CV-qudits and originally designed for the correction of phase distortion induced by self-Kerr effects \cite{SNAPgate}. To inject a group of microwaves into a cavity induces a sum of the phase-rotation gates on each photon-Fock state $\ket{m}$ given by 
\begin{eqnarray}
\label{SNAP01}
\hat{S} = \sum_{m} \exp(i \Phi_m) \ket{m}\bra{m}.
\end{eqnarray}
In our scheme, four groups of microwaves are applied due to $d=4$ to obtain the same phase rotations on each $\ket{k_4}$ ($k=0,1,2,3$) and the grouped phase gate is acheived on each $\ket{{k}_4}$ independently. For example, if we apply the SNAP operation with four-group phase gates, e.g., $\Phi_{4m} = \phi_0$, $\Phi_{4m+1} = \phi_1$, $\Phi_{4m+2} = \phi_2$, and $\Phi_{4m+3} = \phi_3$ on $\ket{\Psi_4}$, the phase-operated qudit is given in
\begin{eqnarray}
\label{SNAP02}
\hat{S} ( \phi_0, \phi_{1} , \phi_{2}, \phi_{3})  \ket{\Psi_4} && =  a e^{i \phi_0} \ket{{0}_4} + b e^{i \phi_{1}} \ket{{1}_4} + c e^{i \phi_{2}} \ket{{2}_4} + d e^{i \phi_{3}} \ket{{3}_4}.
\end{eqnarray}

In particular, we utilise two specific SNAP gates for logical phase gates. The first is a parity-conditional phase gate $\hat{S}^{p1} (\phi) = S ( \phi , -\phi , \phi, -\phi)$ applied to only selected photon states with $ \phi_0 =  \phi_2 =\phi$ and $\phi_1 = \phi_3 = -\phi$. For example, $\hat{S}^{p1} (\phi) \, (\ket{\alpha} \pm \ket{- \alpha} ) = e^{ \pm i \phi} (\ket{\alpha} \pm \ket{- \alpha} )$. 
The other gate is given by $\hat{S}^{p2} (\phi) = \hat{S} ( 0 , 0 , \phi, \pi + \phi)$, which is applied to only selected Fock states with $ \phi_0 =  \phi_1 =0$, $\phi_2 = \phi$, and $\phi_3 = \pi + \phi$. Simple examples are $\hat{S}^{p2} (\phi)\, (\ket{\alpha} + \ket{- \alpha} ) =  (\ket{{0}_4} + e^{ i \phi} \ket{{2}_4}  )$ and $\hat{S}^{p2} (\phi) \, (\ket{\alpha} - \ket{- \alpha} ) =  (\ket{{1}_4} - e^{i \phi} \ket{{3}_4}  )$.
The details of the operations are represented in Section \ref{useful_SNAP}.

\subsection{Logical CV qubit under the presence of photon-loss}
\label{sec:LogicalQubit}
The logical qubits for even photon states are defined in
\begin{eqnarray}
\label{Logical0}
\ket{{0}^L_e} &&= {1\over \sqrt{2}}  (\ket{{0}_4} + \ket{{2}_4})  = N^{+}_{\alpha} (\ket{\alpha} + \ket{-\alpha}) = \ket{SCS^+_{\alpha}}, \\
\label{Logical1}
\ket{{1}^L_e} &&=  {1\over \sqrt{2}} (\ket{{0}_4} - \ket{{2}_4}) = N^{+}_{\alpha} (\ket{i \alpha} + \ket{-i \alpha})= \ket{SCS^+_{i \alpha}},
\end{eqnarray}
where Schr\"odinger cat states are given with $N^{\pm}_{\alpha} = 1/\sqrt{2 (1+e^{-2|\alpha|^2} )}$ in 
\begin{eqnarray}
\label{SCS01}
\ket{SCS^{\pm}_{\alpha}} = N^\pm_{\alpha} \left( \ket{\alpha} \pm \ket{-\alpha}  \right).
\end{eqnarray}
Note that $\ket{{+}^L_e} \equiv \ket{0_4}$ and $\ket{{-}^L_e} \equiv \ket{2_4}$. 
Similarly, for the odd-photon subspace, $\ket{{0}^L_o} =   \ket{SCS^-_{\alpha}}$ and $\ket{{1}^L_o}= -i \ket{SCS^-_{i \alpha}}$.
The two types of logical qubits span only either even- or odd-photon states and a photon-loss error can be monitored and corrected by the real-time parity measurement on the final state \cite{Exp_parity}. 

For example, let us assume that a logical qubit is encoded in $\ket{\Psi^L_e} = a_0 \ket{0^L_e} + a_1 \ket{1^L_e}$, which implies that the information of an arbitrary single qubit can be written in even photon subspace as a logical state. By real-time parity measurements, the cavity state is monitored through a superconducting qubit coupled with a readout resonator. Before cavity photon-loss, the parity measurement  always results in the even state $\ket{\Psi^L_e}$. If the parity changes from even to odd, the updated logical state is equivalent to $\hat{a} \ket{\Psi^L_e} \propto  \ket{\Psi^L_o}  = a_0 \ket{0^L_o} - a_1 \ket{1^L_o}$. Thus, the parity change tells us that the quantum information is preserved against photon-loss but the relative phase is altered.

\subsection{Logical single-qubit gates in a three-qudit cluster state}
\label{Sec:mMBQC}
The essence of MBQC is to create a designed multipartite entangled state initially and to apply sequential measurements on individual qubits will operate one- and two-qubit gates for universal quantum computing \cite{MBQC1, MBQC2}. We now propose a specific protocol to perform a modified MBQC protocol from a three CV-qudit entangled state $\ket{3CS_4}_{ABC}$ given in Eq.~(\ref{3QuditCS01}) 
and its original MBQC from a three-qubit cluster state is described in Section \ref{sec:Original_MBQC}. The CV-qudit measurement schemes are all experimentally viable for logical MBQC using the photon-loss error-correcting code \cite{catcode, Yale-QECC}. 

The first step is to determine the photon parity in the cavity state of the final outcome using the parity measurement on $B$ from Eq.~(\ref{3QuditCS01}). 
Although any alternative implementation of building $\ket{\psi^{ideal} ( {\tau_r / 4} ) }_{ABC}$ is applicable for our initial CV-qudit states (e.g., a scheme in Ref.~\cite{Yale_teleport}), we simply assume that $\ket{\psi^{ideal} ( {\tau_r / 4} ) }_{ABC}$ is initially prepared by a cross-Kerr interaction among the cavities. Then, after the decoupling of all the Kerr-interactions (see Fig.~\ref{fig:02}), the middle cavity state is projected by the parity measurement such as $\hat{P}^{par}_{B} \ket{\psi^{ideal} ( {\tau_r / 4} ) }_{ABC}$ and is given in the even or odd parity state on $B$ such as
\begin{eqnarray}
\label{Special_even01}
\ket{3CS^e_4 }_{ABC}  && = {1\over \sqrt{2}} \Big(  \ket{\tilde{0}_4 }_A \ket{{0}_4}_B \ket{\tilde{0}_4 }_C +  \ket{\tilde{2}_4 }_A \ket{{2}_4}_B \ket{\tilde{2}_4 }_C \Big), ~~~~{\rm (for~even)} \\
\label{Special_even02}
\ket{3CS^o_4 }_{ABC}  && ={1\over  \sqrt{2}} \Big( \ket{\tilde{1}_4}_A \ket{{1}_4}_B \ket{\tilde{1}_4}_C  + \ket{\tilde{3}_4 }_A \ket{{3}_4}_B \ket{\tilde{3}_4 }_C \Big).  ~~~~{\rm (for~odd)}
\end{eqnarray}
Note that this is the only intialisation operation on $B$ to choose the parity of the outcome state, and we do not touch the cavity state in $B$ afterwards.
Without loss of generality, we will assume that the state is subjected in $\ket{3CS^e_4 }_{ABC}$, however, the odd parity case is identical except the definition of logical qubits given in $\ket{{0}^L_o} = \ket{SCS^-_{\alpha}}$ and $\ket{{1}^L_o} = -i \ket{SCS^-_{i \alpha}}$. 

We now perform the cavity operations in $A$ and $C$ with two parameters ($\theta_1$ and $\theta_2$) to obtain desired single-qubit gates on $B$. Because of the lack of cavity measurement capability, we cannot directly perform single-cavity measurement in $\ket{\pm \theta}\bra{\pm \theta}$, however, we alternatively suggest logical single-qubit phase operation first and cavity measurement along the logical $Z$-axis because $\ket{\pm \theta}\bra{\pm \theta} \propto  {R}^Z (-\theta) \ket{\pm}\bra{\pm} ({R}^Z (-\theta))^\dag$. To implement a logical phase gate, SNAP gates are used for encoding the desired operations on logical qubits. More precisely, two SNAP gates, $\hat{S}^{p1} ({\theta_1 / 2})$ on qubit $A$ and $\hat{S}^{p1} (-\theta_1/ 2)$ on $C$, are applied for mimicking a single-qubit phase gate with $\theta_1$. Note that $\hat{S}^{p1}$ is a parity-conditional phase gate as shown in Section \ref{useful_SNAP} and the phase information is embeded in the three CV-qudit state
\begin{eqnarray}
\label{Special_even02-2}
\hspace{-2cm} && \ket{Out_2 (\theta_1)}_{ABC}  = \left( \hat{S}^{p1}_A \left({\theta_1 \over 2} \right) \, \hat{S}^{p1}_C \left(-{\theta_1 \over 2} \right) \right) \ket{3CS^e_4 }_{ABC}, \\
\hspace{-2cm}&& ~~~~~~= {1\over 2} \Big( \left(   \ket{0^L_e}_A \ket{0^L_e}_C  +  \ket{0^L_o}_A   \ket{0^L_o}_C \right) \ket{{0}^L_e} _B +   e^{i\theta_1} \left(  \ket{0^L_e}_A   \ket{0^L_o}_C  +   \ket{0^L_o}_A \ket{0^L_e}_C  \right)  \ket{{1}^L_e}_B  \Big). \nonumber 
\end{eqnarray}
Because the SNAP gates of $\theta_1$ are QND operations, the total cavity stat is not collapsed into a single cavity state yet.

In the next step,  phase $\theta_2$ is imprinted by $\hat{S}^{p2} (\theta_2)$ on $C$ in $\ket{Out_2}_{ABC}$ such as
\begin{eqnarray}
\label{Special_even03}
\hspace{-2.5cm}&&  \ket{Out_3 (\theta_1, \theta_2) }_{ABC}  = \hat{S}^{p2}_C (\theta_2)  \ket{Out_2 (\theta_1)}_{ABC}, \\
\hspace{-2.5cm} &&~~~~~=  {1\over 2 \sqrt{2}} \Big[ \left[ \ket{0^L_e}_A ( \ket{{+}^L_e}_C + e^{i\theta_2} \ket{{-}^L_e}_C )  +  \ket{0^L_o}_A  ( \ket{{+}^L_o}_C - e^{i\theta_2} \ket{{-}^L_o}_C ) \right] \ket{{0}^L_e} _B ,  \nonumber \\
\hspace{-2.5cm} && ~~~~~~~~~ + e^{i\theta_1} \left[  \ket{0^L_e}_A  ( \ket{{+}^L_o}_C - e^{i\theta_2} \ket{{-}^L_o}_C )  + \ket{0^L_o}_A ( \ket{{+}^L_e}_C + e^{i\theta_2} \ket{{-}^L_e}_C ) \right] \ket{{1}^L_e}_B  \Big]. \nonumber
\end{eqnarray}
Although we showed a preferred sequence of SNAP gates performed by $\hat{S}^{p1}$ on $A$ and $C$ first and $\hat{S}^{p2}$ on $C$ second, one can choose an alternative sequence depending on each cavity (e.g., $\hat{S}^{p1}$ on $A$ first and $\hat{S}^{p2}\hat{S}^{p1}$ on $C$ second).

Finally, we are ready to perform the parity and coherent-state measurements on $\ket{Out_3 (\theta_1, \theta_2) }_{ABC}$ to gain the designed local qubit in $B$ given by cavity projections on $A$ and $C$, which is equivalent to the total operation of the original MBQC in Eq.~(\ref{final_qubit}). When we perform the parity measurement on $A$,  the resultant state is equal to $\ket{Out_4 (\theta_1, \theta_2) }_{BC}  \propto \hat{P}^{Par}_A \ket{Out_3}_{ABC}$.
The state for the even parity is given in
\begin{eqnarray}
\label{Special_even03}
\hspace{-2cm} \ket{Out_4^e}_{BC}  &&  = {1\over 2 } \Big[ ( \ket{{+}^L_e}_C + e^{i\theta_2} \ket{{-}^L_e}_C )  \ket{{0}^L_e} _B + e^{i\theta_1}( \ket{{+}^L_o}_C - e^{i\theta_2} \ket{{-}^L_o}_C ) \ket{{1}^L_e}_B  \Big],  
\end{eqnarray}
while the odd one is
\begin{eqnarray}
\label{Special_odd03}
\hspace{-2cm} \ket{Out_4^o}_{BC} &&  = {1\over 2 } \Big[ ( \ket{{+}^L_o}_C - e^{i\theta_2} \ket{{-}^L_o}_C ) \ket{{0}^L_e} _B + e^{i\theta_1}  ( \ket{{+}^L_e}_C + e^{i\theta_2} \ket{{-}^L_e}_C )  \ket{{1}^L_e}_B  \Big].  
\end{eqnarray}

\begin{table}[b]
\centering
\begin{tabular}{|l|l|l|l|}
\hline 
Outcome & Logical gate & Outcome & Logical gate \\ \hline
$\ket{even}_A \ket{\alpha}_C$ & $f_{12} \, {\cal R}^{z} (\theta_1)  {{ \cal H}}\,  { \cal R}^{z} (\theta_2)$ & 
$\ket{odd}_A \ket{\alpha}_C$& $f_{12} {\cal X} {\cal R}^{z} (-\theta_1)  {{ \cal H}}\,  { \cal R}^{z} (\theta_2)$ \\ \hline
$\ket{even}_A \ket{-\alpha}_C$ & $f_{12} \, {\cal Z} {\cal R}^{z} (\theta_1)  {{ \cal H}}\,  { \cal R}^{z} (\theta_2)$ &
$\ket{odd}_A \ket{-\alpha}_C$ & $f_{12} {\cal X Z} {\cal R}^{z} (-\theta_1)  {{ \cal H}}\,  { \cal R}^{z} (\theta_2)$  \\ \hline
$\ket{even}_A \ket{i\alpha}_C$ & $f\rq{}_{12} {\cal Z} {\cal X}  {\cal R}^{z} (-\theta_1)  {{ \cal H}}\,  { \cal R}^{z} (\theta_2)$ &
$\ket{odd}_A \ket{i\alpha}_C$ & $f\rq{}\rq{}_{12} {\cal R}^{z} (\theta_1)  {{ \cal H}}\,  { \cal R}^{z} (\theta_2)$ \\ \hline
$\ket{even}_A \ket{-i\alpha}_C$ & $f\rq{}_{12}  {\cal X}  {\cal R}^{z} (-\theta_1)  {{ \cal H}}\,  { \cal R}^{z} (\theta_2)$ &
$\ket{odd}_A \ket{-i\alpha}_C$&  $f\rq{}\rq{}_{12} e^{i \pi } {\cal Z}  {\cal R}^{z} (\theta_1)  {{ \cal H}}\,  { \cal R}^{z} (\theta_2)$  \\ \hline
\end{tabular}
\caption{Table for measurement outcomes in $A$ and $C$ and the performed logical single-qubit gates ($f\rq{}_{12} = f_{12} \, e^{i \frac{ \pi }{4}  }\,  {\cal R}^{z} ({\pi \over 2})$ and $f\rq{}\rq{}_{12} =f_{12} \, e^{-i \frac{ \pi }{4}  }\,  {\cal R}^{z} ({\pi \over 2})$).}
\label{Table2}
\end{table}

Then, if we project the qubit $C$ by the coherent state-measurement $\{\ket{\alpha}, \ket{i\alpha}, \ket{-\alpha}, \ket{-i\alpha}\}$ as shown in Section \ref{sec:coherent-measurement}, the successful detection gives the logical qubit in
\begin{eqnarray}
\label{Special_even04}
&& \hspace{-1.5cm} \ket{Out_5^{\alpha,e}}_{B} = \sqrt{2} \, {}_{C} \langle \alpha \ket{Out_4^{e}}_{BC}  = e^{\frac{i }{2} (\theta_1+\theta_2) } {\cal R}^{z} (\theta_1)  {{ \cal H}}\,  { \cal R}^{z} (\theta_2)  \ket{{+}^L_e}_B, \\
&&\hspace{-1.5cm} \ket{Out_5^{-\alpha,e}}_{B} =e^{\frac{i }{2} (\theta_1+\theta_2) } {\cal Z} {\cal R}^{z} (\theta_1)  {{ \cal H}}\,  { \cal R}^{z} (\theta_2)  \ket{{+}^L_e}_B, \\
&&\hspace{-1.5cm} \ket{Out_5^{i\alpha,e}}_{B} =e^{\frac{i }{2} (\theta_1+\theta_2+{\pi \over 2}) } {\cal R}^{z} ({\pi \over 2}) {\cal Z} {\cal X}  {\cal R}^{z} (-\theta_1)  {{ \cal H}}\,  { \cal R}^{z} (\theta_2)  \ket{{+}^L_e}_B,\\
&& \hspace{-1.5cm} \ket{Out_5^{-i\alpha,e}}_{B} =e^{\frac{i }{2} (\theta_1+\theta_2+{\pi \over 2}) }{\cal R}^{z} ({\pi \over 2})  {\cal X}  {\cal R}^{z} (-\theta_1)  {{ \cal H}}\,  { \cal R}^{z} (\theta_2)  \ket{{+}^L_e}_B, 
\end{eqnarray}
and
\begin{eqnarray}
\label{Special_even05}
&& \hspace{-1.5cm} \ket{Out_5^{\alpha,o}}_{B} = e^{\frac{i }{2} (\theta_1+\theta_2) } {\cal X} {\cal R}^{z} (-\theta_1)  {{ \cal H}}\,  { \cal R}^{z} (\theta_2)  \ket{{+}^L_e}_B,  \\
&& \hspace{-1.5cm} \ket{Out_5^{-\alpha,o}}_{B} = e^{\frac{i }{2} (\theta_1+\theta_2) } {\cal X} {\cal Z}  {\cal R}^{z} (-\theta_1)  {{ \cal H}}\,  { \cal R}^{z} (\theta_2)  \ket{{+}^L_e}_B, \\
&& \hspace{-1.5cm} \ket{Out_5^{i\alpha,o}}_{B} =e^{\frac{i }{2} (\theta_1+\theta_2- {\pi \over 2}) }  {\cal R}^{z} ({\pi\over 2}) {\cal R}^{z} (\theta_1)  {{ \cal H}}\,  { \cal R}^{z} (\theta_2)  \ket{{+}^L_e}_B , \\
&& \hspace{-1.5cm} \ket{Out_5^{-i\alpha,o}}_{B} = e^{\frac{i }{2} (\theta_1+\theta_2+ 3{\pi \over 2}) } {\cal R}^{z} ({\pi \over 2}) {\cal Z} {\cal R}^{z} (\theta_1)  {{ \cal H}}\,  { \cal R}^{z} (\theta_2)  \ket{{+}^L_e}_B, 
\end{eqnarray}
where ${ \cal R}^{z}$ and ${{ \cal H}}$ are a logical rotation gate on z-axis and a logical Hadamard gate defined by logical qubits in $\ket{{0}^L}$ and $\ket{{1}^L}$. Note that a repeat-until-success method can be used for approximated orthogonal projection of the cavity states on the measurement set of $\{ \ket{\alpha}\bra{\alpha}, \ket{i\alpha}\bra{i\alpha}, \ket{-\alpha}\bra{-\alpha}, \ket{-i\alpha}\bra{-i\alpha} \}$ for large $\alpha \ge 2$. 

The details of logical gates with respect to each outcome are presented in Table \ref{Table2}.
Note that logical Pauli operators ${\cal Z} \equiv \left( \hat{X}_4 \right)^2$ and  ${\cal X} \equiv \hat{Z}_4$ can be defined by CV-qudit Pauli gates in Section \ref{2-1}.
Therefore, it is shown that the specific logical operation of mMBQC is performed by the sequential operations and measurements in the cavities of $A$ and $C$.

\subsection{Implementation of a two CV-qudit state in the JC Hamiltonian} 
\label{Implement_s02} 

We here mainly examine how to build two-qudit entangled states in the model of the JC generalised Hamiltonian ($\hat{H}^{JC}_{ABM} $), which describes the nonlinear effects given from the contribution of the intermediary transmon qubit (upto the third level). From the JC Hamiltonian in Eq.~(\ref{total_Ham01}) with two coherent states, the total state in the two cavities with the qubit evolves in time and the state of two cavities are given by
\begin{eqnarray}
\label{Time-evol-sim02}
\ket{\psi^{JC} (t)}_{ABM} = \exp \left( i \, \hat{H}^{JC}_{ABM} \, t \right) \ket{\alpha}_{A} \ket{\alpha}_{B} \ket{g}_{M} , \\
\label{Time-evol-sim03}
\rho^{JC}_{AB} (t) = Tr_{M} \Big( \ket{\psi^{JC} (t)}_{ABM} \bra{\psi^{JC} (t)} \Big).
\end{eqnarray}
\begin{figure}[t]
\includegraphics[width=14cm,trim= 0cm 0cm 2cm  0cm] {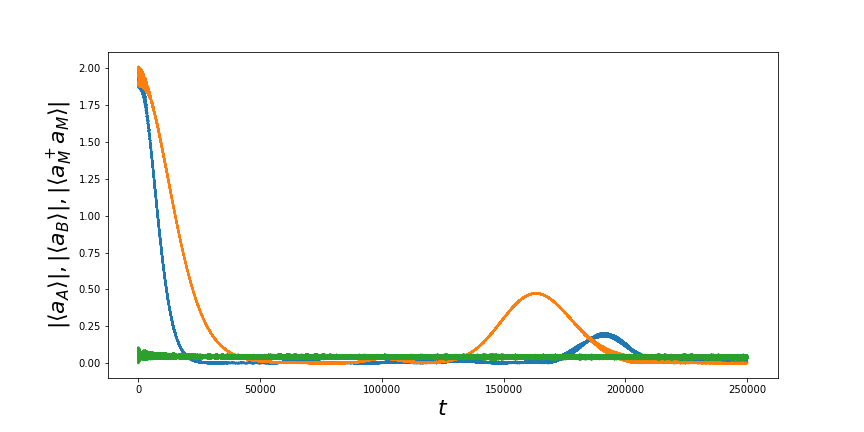}
\center
\includegraphics[width=12cm,trim= 3cm 0cm 2cm  0cm]{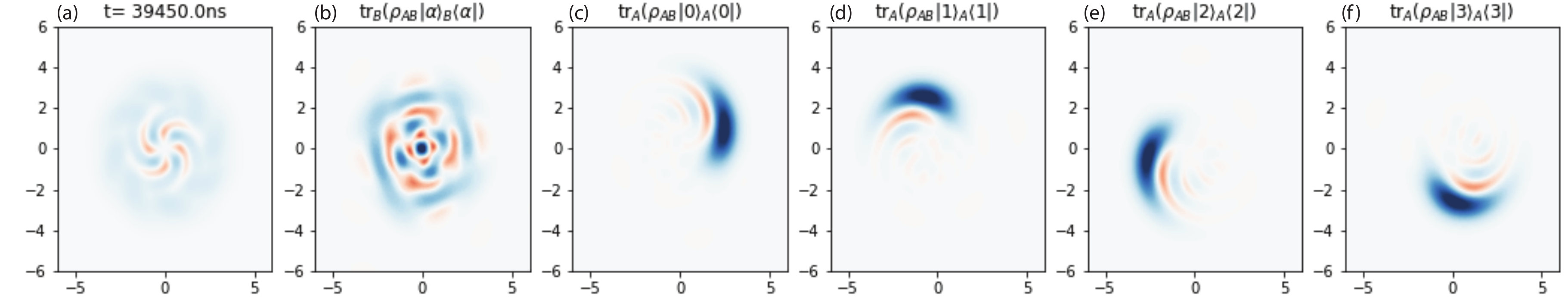}
%20180620_Wigner_Km06_at39450.png}
\caption{(Top) $|\langle a_A \rangle |$,  $|\langle a_B \rangle |$ and $|\langle a^\dag_M a_M \rangle|$ are shown from the initial state $\ket{\alpha}_{A} \ket{\alpha}_{B}$ ($\alpha = 2.0$) under the generalised JC Hamiltonian in Eq.~(\ref{total_Ham01}) with the parameters of $\omega_A = 5.5$ GHz, $\omega_B = 8.5$ GHz, $\omega_{M} = 4.0$ GHz, $\lambda_{AM} = 0.12$ GHz, $\lambda_{BM} = 0.15$ GHz and  $K_M = - 0.6$  GHz. While the orange line shows the revival of mode $B$ at $\tau_r \approx 160\,\mu$s, the green line indicates the expectation value $|\langle a^\dag_M a_M \rangle| \approx 0$, which shows the ground state $\ket{g}_M$ mostly as predicted in the adiabatic method.
(Bottom) In (a), a mixture of four coherent states is given by $tr_B \Big( \rho^{JC}_{AB} (t_0) \Big)$ at $t_0 =39.45\,\mu$s ($\approx \tau_r /4$) while the Wigner plot in (b) indicates that the evolved state $ \rho^{JC}_A = \left( {}_{B} \bra{\alpha} \rho^{JC}_{AB} (t_0)\ket{\alpha}_{B} \right)$ is also very close to the state $\ket{{0}_4}_A$ with $F\approx0.978$. From (c) to (f), we project the state on the Fock states from $\ket{0}_A\bra{0}$ to $\ket{3}_A\bra{3}$ and the Wigner plots of $ \rho^{JC, k}_B$ are shown as coherent states in $\ket{\tilde{k}_4}$ ($k=0,1,2,3$). 
}
\label{fig:03}
\end{figure}

In Fig.~\ref{fig:03}, we numerically illustrate the dynamics of cavity states evolved by the JC Hamiltonian $\hat{H}^{JC}_{ABM}$ to create the two CV-qudit cluster state given in Eq.~(\ref{Simple_02}). The realistic parameters are chosen in $\omega_A = 5.5$ GHz, $\omega_B = 8.5$ GHz, $\omega_{M} = 4.0$ GHz, $\lambda_{AM} = 0.12$ GHz, $\lambda_{BM} = 0.15$ GHz and  $K_M = - 0.6$ GHz. In the top of Fig.~\ref{fig:03}, the revival picks appear at around $t \approx 160 \mu$s with $\alpha = 2.0$ as given in the values of $|\langle a_A \rangle |$ (blue) and  $|\langle a_B \rangle |$ (orange). Note that $|\langle a  (t) \rangle | =0$ implies that the cavity states are the evenly distributed coherent states in phase space while $|\langle a^\dag_M a_M \rangle| \approx 0$ does that the transmon qubit is almost nearly in $\ket{g}_M$ such that $\ket{\psi^{JC} (t)}_{ABM} \bra{\psi^{JC} (t)}\approx \rho^{JC}_{AB} ({t} )  \otimes \ket{g}_M\bra{g}$.

What we would like to find is that the state $\rho^{JC}_{AB} (t_0)  \approx \ket{\psi^{ideal}  ( {\tau_r / 4} ) }_{AB} \bra{\psi^{ideal}  ( {\tau_r / 4} ) }$ at certain time $t_0$ (see the details in Eq.~(\ref{Simple_02})). To compare $\rho^{JC}_{AB} (t_0)$ with the ideal two-qudit state (given in Eq.~(\ref{ideal2qudit}), one may obtain the fidelity between the two states, however, this value might not represent the characteristics of the time-evolved state $\rho^{JC}_{AB} (t_0)$ because the distortion of the cavity state from the self-Kerr effects suppress the fidelity very low.
In the spirit of MBQC, one of the simple verifications of the measured states is to compare between the projected cavity states of $\rho^{JC}_{AB} $ and  of $\ket{\psi^{ieal} }_{AB}$. In Fig.~\ref{fig:03}(a), the state in $A$ is given by $tr_{B} \left( \rho^{JC}_{AB} (t_0) \right)$, in which we expect to obtain the mixture of four coherent states at $t_0=39.45\,\mu$s. From ($b$) to ($f$), we plot the Wigner functions of the cavity state in mode $A$ ($B$) at $t_0$ given by the projection of the certain states in mode $B$ ($A$) such as
\begin{eqnarray}
\label{Outcome-sim01}
\hspace{-1cm} \rho^{JC}_A \propto tr_B \Big( \left( \openone_A \otimes \ket{\alpha}_{B} \bra{ \alpha}\right) \rho^{JC}_{AB} ( t_0 )  \Big) \approx \ket{{0}_4}_A \bra{{0}_4}, \\
\hspace{-1cm} \rho^{JC, k}_B \propto tr_A \Big(( \ket{k}_{A} \bra{ k} \otimes \openone_B )\, \rho^{JC}_{AB} (t_0) \Big)  \approx \ket{\tilde{k}_4}_B \bra{\tilde{k}_4}  = \ket{\alpha e^{i k {\pi / 2}}}_B  \bra{\alpha e^{i k {\pi / 2}}},
\end{eqnarray}
for $k=0,1,2,3$. In the bottom of Fig.~\ref{fig:03}, we show that the maximum fidelity $F=  \left| {}_A \bra{{0}_4}  \rho^{JC}_A \ket{{0}_4}_A  \right|$ is approximately 0.978 at $t_0 \approx 40\,\mu$s in (b) and some levels of self-Kerr distortions occur during the time evolution from (c) to (f). We neglect decoherence processes in the cavities since the state-of-the-art lifetime of a 3D cavity is above 1.2 ms and the decoherence is expected to be not dominant until the period $t_0 \approx 40\,\mu$s. Apparently, this period of creating multi-partite microwave entangled state could not grow up much with increasing the number of cavities.

\section{Conclusion}
In summary, we introduce a new-type of CV logical MBQC in three microwave cavities coupled with superconducting qubits in a circuit-QED system. After the CV-qudits are defined, three specific circuit-QED gates are introduced to realise logical gate operations for the protocol of logical MBQC. We deliver the method of a logical single-qubit gate in photon-loss correcting codes from the three CV-qudit entangled state. Finally, the implementation of the two CV-qudit state and measured cavity states are numerically investigated under the JC Hamiltonian in a two-cavity system coupled with a superconducting qubit. The results show that the entangled CV-qudit states can be efficiently built with high fidelity (above 0.97) via the cross-Kerr effect induced by the intermediary superconducting qubit between cavities. 

%{\bf To include the third level of the transmon makes a change of the revival time longer ($\approx XXX~\mu$s), which indicates that the cross-Kerr effect is reduced, however the fidelity is roughly the same as the two-level system in a transmon qubit. 
%Because the state-of-the-art lifetime of a 3D cavity is above 1.2 ms, the decoherence might be not dominant during the the period of creating the desired states even in the three-level case.

\section{Methods}
\subsection{How to build two CV-qudit states}
\label{Sec:Entangle-two-cat}
When an initial state $\ket{\psi^{int}}_{AB}$ is prepared in cavities $A$ and $B$, the time-evolved state at time $t$ is given by 
\begin{eqnarray}
\label{total_Ham00}
\ket{\psi (t)}_{AB} = \exp \left( i \hat{H}_{AB}  \, t \right) \ket{\psi^{int}}_{AB},
\end{eqnarray}
where the cross-Kerr Hamiltonian is $\hat{H}_{AB} = K_{AB} (\hat{a}^{\dag}_A \hat{a}_A) (\hat{a}^{\dag}_B \hat{a}_B)$ and $K_{AB}$ is the strength of cross-Kerr interaction. 
The initial state is fully revived at $t = \tau_r =  {2 \pi / \,K_{AB}}$, and the evolved state is in general written in an entangled (inseparable) state between two modes at $t \neq \tau_r $. For $t = \tau_r /d$, it is given by
\begin{eqnarray}
\label{Simple_00}
\ket{\psi^{ideal} ( {\tau_r / d} ) } = \exp \left( i {2 \pi \over d} (\hat{a}^{\dag}_A \hat{a}_A) (\hat{a}^{\dag}_B \hat{a}_B)  \right) \ket{\psi^{int}}_{AB}.
\end{eqnarray}
For example, for $t =  {\tau_r / 2} $ with $\ket{\psi^{int}}_{AB} = \ket{\alpha}_{A} \ket{\alpha}_{B}$, the state evolves such as
\begin{eqnarray}
\label{Simple_01}
\ket{\psi^{ideal} ( {\tau_r / 2} ) }_{AB} &&= {1 \over \sqrt{2}} \left( \ket{SCS^{+}_{\alpha}}_A  \ket{\alpha }_B + \ket{SCS^{-}_{\alpha}}_A \ket{-\alpha }_B \right).
\end{eqnarray}

This state is known as an entangled coherent state \cite{Sanders12, Sanders92}, which is also of excellence for quantum metrology and other QI processing methods \cite{JooPRL11, Ralph03} and has been recently demonstrated in a deterministic method in circuit-QED \cite{Yale_ECS} and probabilistically in quantum optics \cite{Paris_ECS}.
In fact, the entangled coherent state can be used as a simplest resource state for MBQC with no error-correction because CV quantum teleportation, which is the building block for MBQC, has been demonstrated in quantum optics \cite{Vaidman, compare_CV_tele1, CV_RMP2} and investigated in circuit-QED \cite{Joo_SR}. The similar method of implementing the states has been suggested with the assumption of the cross-Kerr interaction in a circuit-QED system \cite{newarxiv_ECS}.

For $d=4$, the time evolution time is the half period of $\ket{\psi^{ideal} ( {\tau_r / 2} ) }_{AB} $. The evolved state at $t =  {\tau_r / 4} $ is written by 
\begin{eqnarray}
\label{Simple_02}
\hspace{-1cm} \ket{\psi^{ideal}  ( {\tau_r / 4} ) }_{AB} && = {1\over 2} \Big( \ket{{0}_4}_A \ket{\alpha }_B +\ket{{1}_4}_A \ket{i \alpha }_B + \ket{{2}_4}_A \ket{-\alpha }_B +\ket{{3}_4}_A \ket{-i\alpha }_B \Big),
\nonumber \\ 
&&=  {1 \over 2}  \sum_{k=0}^3 \ket{{k}_4}_A \ket{\tilde{k}_4}_B.  \label{ideal2qudit}
 \end{eqnarray}
This state $\ket{\psi^{ideal}  ( {\tau_r / 4} ) }_{AB}$ is a CV version of a two-qudit cluster state. Alternatively, the equivalent CV-qudit state has been very recently realised for qudit quantum teleportation \cite{Yale_teleport}.

\subsection{Fock- and coherent-state projections on a cavity state}
\label{sec:coherent-measurement}
One of the important techniques in circuit-QED is based on a conditional qubit-rotation depending on a chosen Fock state $\ket{m}_A$ and the projection measurement set is given by $\hat{P}^{Foc} (m) = \{ \ket{m}\bra{m}, \openone - \ket{m}\bra{m} \}$ \cite{BlaisPRA}. 
For example, let us assume that a coherent state $\ket{\alpha}$ is prepared in cavity $A$ with the ground state $\ket{g}$ such as $\ket{\alpha}_A \ket{g}_{J} = \sum_{m} c_m \ket{m}_{A} \ket{g}_{J}$ for $c_m = \bra{m} \alpha \rangle$. 
A conditional qubit-rotation gate is effectively performed on photon state $\ket{m}$ represented by 
\begin{eqnarray}
\label{total_StateProj01}
\hat{R}^{y}_{A\, J} (m, \phi) = \sum_{n \ne m} e^{i \eta_n} \ket{n}_{A}\bra{n} \otimes \openone_{J}  + \ket{m}_{A}\bra{m} \otimes \hat{R}^y_{J} (\phi),
\end{eqnarray}
where $\hat{R}^y (\phi) = \cos {\phi\over 2} \, \openone - i \sin {\phi\over 2} \, Y = \left(
\begin{array}{cc}
 \cos {\phi\over 2} & - \sin {\phi\over 2} \\
 \sin {\phi\over 2} & \cos {\phi\over 2} \\
\end{array}
\right)$. 

For $\phi = {\pi }$, the state becomes
$\hat{R}^{y}_{A\, J} (m, {\pi }) \, \ket{\alpha}_A \ket{g}_{J} =  \sum_{n \neq m} c_n e^{i \eta_n} \ket{n}_{A} \ket{g}_{J} + c_m \ket{m}_{A} \ket{e}_{J}$ where $ e^{i \eta_n}$ is an undesired operation in $\hat{R}^{y}_{A\, J} (m, \pi)$ due to self-Kerr interaction but does not influence our result because we only use the outcome state $\ket{e}$ in a heralded way \cite{Yale_big_cat}.
Then, when the outcome is measured in $\ket{e}_{J}$, the cavity state is also projected in $\ket{m}_A$ and the operator of this Fock-state projection on the $m$-th photon is given by
\begin{eqnarray}
\label{total_StateProj02}
\hat{P}^{Foc}_{A} (m) = \left( \openone_{A} \otimes \ket{e}_{J} \bra{e} \right) \, \hat{R}^{y}_{A\,J} (m, \pi).
\end{eqnarray}
In the unsuccessful case of measurement in $\ket{g}$, the cavity state is projected by the operator $\left( \openone_A - \ket{m}_A \bra{m} \right) $ and we can perform the repeat-until-success protocol $\hat{P}^{Foc}_{A} (p)$ for $p \neq m$. 

A coherent-state projection can be also performed by adding displacement operation $\hat{D}^{-\alpha} = e^{\alpha^* a - \alpha a^{\dag}}$ on cavity states \cite{Yale_big_cat}. 
the coherent-state projection on $\ket{\alpha}$ is given in
\begin{eqnarray}
\label{total_StateProj03}
\hspace{-1.5cm} \hat{P}^{Coh}_{A} (\alpha) = \hat{P}^{Foc}_{A} (0) \left( \hat{D}^{-\alpha}_A \otimes  \openone_{J} \right) 
=  \left( \openone_{A} \otimes \ket{e}_{J} \bra{e} \right) \, \hat{R}^{y}_{A\,J} (0, \pi) \, \left( \hat{D}^{-\alpha}_A \otimes  \openone_{J} \right).
\end{eqnarray} 

\subsection{Parity measurement on a cavity state}
\label{sec:parity}
When we first perform the operation $\hat{R}^y_J (\pi/2) $ on the initial transmon state $\ket{g}_{J}$, a conditional cavity-rotation gate $\hat{C}^{p} (\varphi)$ is given by
\begin{eqnarray}
\label{arbit_CVqudit02}
\hat{C}^{p}_{A\,J} (\varphi) \left[ {1\over \sqrt{2} }  \ket{\alpha}_A (\ket{g}_{J} + \ket{e}_{J}) \right] =   {1\over \sqrt{2} }  \left( \ket{\alpha}_A \ket{g}_{J} + \ket{\alpha\ e^{i \varphi} }_A  \ket{e}_{J} \right)  ,
\end{eqnarray}
and the operated state with $\varphi=\pi$ is represented by 
\begin{eqnarray}
\label{arbit_CVqudit03}
\hspace{-1cm}
\hat{C}^{p}_{A\,J} (\pi) \, \hat{R}^y_{J} (\pi/2) \, \ket{\Psi_4}_A \ket{g}_{J}  &=& \ket{\Psi_4}_A \ket{g}_{J} \nonumber \\
&& + \left( a \ket{{0}_4}_A - b \ket{{1}_4}_A + c \ket{{2}_4}_A - d \ket{{3}_4}_A\right) \ket{e}_{J}.
\end{eqnarray}

Finally, if we apply an additional $\hat{R}^y_{J} (\pi/2)$ and measure the superconducting qubit in $\{ \ket{g}_{J}\bra{g}, \ket{e}_{J}\bra{e} \}$, the cavity state is projected on the even- or odd-photon subspace such as parity measurement $\hat{P}^{Par} (even/odd) = \{ \ket{{0}_4} \bra{{0}_4} + \ket{{2}_4} \bra{{2}_4} ,  \ket{{1}_4} \bra{{1}_4}+ \ket{{3}_4} \bra{{3}_4} \}.$
For example, if the superconducting qubit is measured in $\ket{e}$ (or $\ket{g}$), the total state is projected in even (odd) photon numbers and the parity measurement is represented by 
\begin{eqnarray}
\label{Parity02}
\hat{P}^{Par} (even/odd) = \ket{e/g}_{J} \bra{e/g}\, \hat{R}^y_{J} (\pi/2) \, \hat{C}^{p}_{A\,J} (\pi) \, \hat{R}^y_{J} (\pi/2).
\end{eqnarray}

\subsection{SNAP gate for a logical single-qudit phase gate}
\label{useful_SNAP} 
The original motivation of SNAP gate was to cancel out the self-Kerr defect in each cavity independently because self-Kerr effects dominantly influence the shape of the cavity state in a physical setup if the evolution time is not short. This unique circuit-QED technique works in a dispersively coupled cavity-transmon system \cite{SNAPgate} and has been demonstrated to minimize phase distortions acquired during the self-Kerr interaction period. The dispersive energy shifts of the cavity system allow a phase gate in individual Fock states to be addressed by driven microwaves. 

For $\hat{S}^{p1}$, the outcome state from $\ket{\Psi_4}$ is given by
a grouped phase gate dependent on photon parities such as 
\begin{eqnarray}
\label{SNAP03}
\hat{S}^{p1} (\phi)  \ket{\Psi_4}  && =  e^{i \phi} \left( a \ket{{0}_4}  + c \ket{{2}_4} \right) + e^{-i \phi} \left( b \ket{{1}_4}  + d \ket{{3}_4} \right),
\end{eqnarray}
while that for $\hat{S}^{p2}$
\begin{eqnarray}
\label{SNAP04}
\hat{S}^{p2} (\phi)  \ket{\Psi_4}  && = a \ket{{0}_4} + b \ket{{1}_4} + e^{ i \phi} c \ket{{2}_4} - d  e^{ i \phi} \ket{{3}_4}. 
\end{eqnarray}

\begin{table}[b]
\centering
\begin{tabular}{|c|c|}
\hline 
Outcome state in $C$ & Single-qubit operations  \\ \hline
$\ket{Out^{+_1+_2} (\theta_1, \theta_2) }_{C}$ &  $ f_{12} R^z (\theta_1)  H\,  R^z (\theta_2)  $ \\ \hline
$\ket{Out^{-_1+_2} (\theta_1, \theta_2) }_{C}$ & $f_{12} \,Z\, R^z (\theta_1)  H\,  R^z (\theta_2)  $ \\ \hline
$\ket{Out^{+_1-_2} (\theta_1, \theta_2) }_{C}$ & $f_{12} \,X\, R^z (-\theta_1)  H\,  R^z (\theta_2)$ \\ \hline
$\ket{Out^{-_1 -_2} (\theta_1, \theta_2) }_{C}$ & $f_{12} \,ZX\, R^z (-\theta_1)  H\,  R^z (\theta_2) $ \\ \hline
\end{tabular}
\caption{Table for outcomes and performed gates in Section \ref{sec:Original_MBQC} ( $f_{12} =e^{\frac{i }{2} (\theta_1+\theta_2) }$).}
\label{Table1}
\end{table}

\subsection{MBQC in a three-qubit cluster state}
\label{sec:Original_MBQC}
We here describe the original MBQC protocol in a three-qubit cluster state.
If three qubits are initially prepared in $\ket{+}$ in $A$, $B$, and $C$, two CZ gates between $A$ and $B$ as well as $B$ and $C$, which construct a three-qubit cluster state given in
\begin{eqnarray}
\label{GHZ_01}
\ket{3CS}_{ABC}  && = {1\over \sqrt{2}} (  \ket{0 }_A \ket{+}_B \ket{0 }_C +  \ket{1 }_A \ket{-}_B \ket{1 }_C ).
\end{eqnarray}
In the frame of MBQC, qubits are sequentially measured in the basis vectors of $ \ket{\pm \theta} = (\ket{0} \pm e^{-i \theta} \ket{1})/\sqrt{2}$. For example, if $\ket{\pm \theta_1}$ is measured in qubit $A$ in Eq.~(\ref{GHZ_01}), the resultant state is  given by
\begin{eqnarray}
\label{GHZ_02}
\hspace{-2.5cm}
\ket{Out^{\pm}}_{BC} &&= \sqrt{2}\, {}_{A} \langle \pm \theta_1 \ket{3CS}_{ABC}= {1\over 2} \left[  \ket{0}_B ( \ket{0 }_C \pm  e^{i \theta_1} \ket{1 }_C) + \ket{1}_B ( \ket{0 }_C \mp  e^{i \theta_1} \ket{1 }_C)  \right].
\end{eqnarray} 
In the case that the outcome is $\ket{+\theta_2}_{B}$, the final outcome state is equal to
\begin{eqnarray}
\hspace{-2cm} 
\ket{Out^{+_1 +_2} (\theta_1, \theta_2) }_{C} && =  \sqrt{2} \,{}_{B} \langle +\theta_2 \ket{Out^+ (\theta_1) }_{BC}= e^{\frac{i }{2} (\theta_1+\theta_2) } R^z (\theta_1)  H\,  R^z (\theta_2)  \ket{+}_C. \label{final_qubit}
\end{eqnarray}
As shown in Table \ref{Table1}, this protocol is equivalent to two single-qubit rotations and two sequential projective measurements on $A$ and $B$. Thus, this procedure of MBQC is equivalent to the operation of two single-qubit gates with phases $\theta_1$ and $\theta_2$.

\section{Acknowledgements}
JJ thanks D. Jaksch, E. Ginossar, Y. Nakamura and M. Elliott for fruitful discussions. This work was supported by the KIST Institutional Program (Project No. 2E26680-16-P025) and EPSRC National Quantum Technology Hub in Networked Quantum Information Technology (EP/M013243/1).

\section{Author Contributions}
J.J. and J. K. proposed and devised the basic model. J. J and C.-W. L. performed the analytic and numerical calculations under the guidance of S. K. and Y. N. for experimental viewpoints. All the authors discussed the results and contributed to the preparation of the manuscript.

\end{document}